\documentclass[aps,showpacs,twocolumn,amsmath,preprintnumbers,nofootinbib,a4paper,secnumarabic]{revtex4} 

\usepackage[utf8]{inputenc}
\usepackage{color}\def\rd#1{\textcolor{red}{#1}}
\usepackage{graphicx,epsfig,verbatim}
\usepackage{amssymb,amsfonts}

\usepackage[caption=false]{subfig}
\def\sct{0.51} 
\def\sct{0.49} 
\def\sctt{0.48} 

\def\art{paper}

\def\jrn#1#2#3#4#5#6{\textit{#3} \textbf{#4}, #5 (#6).}   \def\andd{, and } \def\eq{Eq. }  \def\eqs{Eqs. } \def\Ref{Ref. } \def\Refs{Refs. } \def\fig{Fig. } \def\figs{Figs. }

\def\scn#1#2{\section{#1}\lb{#2}}

\def\bfl{\begin{flushleft}}
\def\efl{\end{flushleft}}
\def\bfr{\begin{flushright}}
\def\efr{\end{flushright}}
\def\bc{\begin{center}}
\def\ec{\end{center}}
\def\be{\begin{equation}}
\def\ee{\end{equation}}
\def\bse{\begin{subequations}}
\def\ese{\end{subequations}}
\def\ba{\begin{eqnarray}}
\def\ea{\end{eqnarray}}
\def\baa#1{\begin{array}{#1}}
\def\eaa{\end{array}}
\def\bw{\begin{widetext}}
\def\ew{\end{widetext}}
\def\nn{\nonumber }
\def\lb#1{\label{#1}}
\def\bit{\begin{itemize}}
\def\eit{\end{itemize}}
\def\bco{}
\def\bcs{\begin{cases}}
\def\ecs{\end{cases}}

\def\lan{{\cal L}}
\def\lanp{V} 

\def\drm{d}
\def\dvol{\drm^3 \vec x}

\def\dn{\rho}  \def\dnc{\bar{\dn}}
\def\tps{T_\Psi}

\def\massf{{\cal M}}
\def\kspl{\tilde{k}}
\def\sigc{\sigma_0}
\def\sspl{\tilde\sigma_0}

\begin{document}

\preprint{\small \footnotesize Low Temp. Phys. \textbf{47}, 89 (2021)   
\ \ 
[\href{https://doi.org/10.1063/10.0003166}{DOI: 10.1063/10.0003166}]
}

\title{
Superfluid stars and Q-balls in curved spacetime  
}

\author{Konstantin G. Zloshchastiev}
\email{https://bit.do/kgz}
\affiliation{Institute of Systems Science, Durban University of Technology, Durban 4000, South Africa}

\begin{abstract}
Within the framework of the theory of strongly-interacting quantum Bose liquids, we consider a general relativistic model of self-interacting complex scalar fields with logarithmic nonlinearity taken from dense superfluid models.
We demonstrate the existence of gravitational equilibria in this model, 
described by spherically symmetric nonsingular finite-mass asymptotically-flat solutions.
These equilibrium configurations can describe both massive astronomical objects, such as bosonized superfluid stars or cores of neutron stars, and finite-size particles and non-topological solitons, such as Q-balls.
We give an estimate for masses and sizes of such objects.
\end{abstract}

\date{received: 26 May 2020 [APS], 3 September 2020 [FNT]}


\maketitle

\scn{Introduction}{s:intro}
In the hierarchy of superdense stars,
various objects exist, which occupy an intermediate place between neutron stars and black holes.
To a distant observer, such objects would look almost like black holes; but they have no horizon, for which reason they are often aggregated under the name of compact stars (CS) and black hole mimickers (BHM).
Widely known examples of such objects include geons, quark stars, gravastars; and the
most popular of them, boson stars, which have long since been studied \cite{fi48,ka68,rb69,ty84,csw86}.
In this \art, we propose another type of CS/BHM-type bosonic objects -- superfluid stars,
which are modeled by scalar field with logarithmic nonlinearity
motivated by the theory of strongly-interacting dense
superfluids.
Furthermore, we demonstrate that our model can also describe composite particle-like objects, or Q-balls,
which self-interact and curve spacetime.

Because superfluids are macroscopic wave-mechanical objects, the stability of superfluid stars against gravitational collapse is expected to be enhanced by the uncertainty principle, similar to  boson star models.
Moreover, superfluidity introduces an additional effect here.
The
inviscid flow, caused by suppression of dissipative fluctuations, makes the fluid parcels and 
volume elements
more resistant to coming to a full stop and adhering to each other. 
Therefore superfluid stars are expected to have a larger degree of resistance to gravitational collapse than the conventional boson stars. 

Logarithmic Bose liquids are nonlinear effective models, which have successfully been
used to describe laboratory superfluids, such as the helium II phase \cite{z12eb,sz19,z19ijmpb}.
Unlike models
with quartic nonlinearity, such as the Gross-Pitaevskii one,
logarithmic models go well beyond the two-body interaction approximation. 
In fact, logarithmic nonlinearity occurs in a leading-order approximation of any strongly-interacting 
many-body systems; which can be described in terms of collective degrees of freedom by a single wavefunction, and for which  typical interparticle interaction potentials are much larger than kinetic energies \cite{z18zna}.
This nonlinearity also occurs in a theory of superflow-induced spacetime and emergent gravity \cite{z10gc,z11appb,z20un1}.

While the original logarithmic models are non-relativistic, their Lorentz-symmetric analogues are straightforward to construct.
Relativistic logarithmic scalar fields are known to possess a dilatation symmetry \cite{ros69},
which makes them universally usable over a large range of length and mass scales.
This feature, together with the superfluid-enhanced stability against gravitational collapse, could result in the substantial
increase of the maximal mass of logarithmic superfluid stars,
which is a conjecture to be checked, among other things, in this \art.

\scn{The model}{s:mod}
Adopting the units $c=1$ and metric signature $(-+++)$,
we write a classical Lagrangian of the model:
\be\lb{elagr}
\lan = \frac{R}{16\pi G} -
      \frac{1}{2}\nabla_\mu \phi^* \nabla^\mu
        \phi - 
\lanp (\phi, \phi^*)
,
\ee
where 
the potential of
a minimally coupled complex scalar field $\phi$
is given by
\be\lb{epot}
\lanp (\phi, \phi^*)
= 
- b \left| \phi \right|^2 \left[  
    \ln \left( 
    	\left| \phi \right|^2/a 
    \right) - 1 
    \right]
,
\ee
where $a$ and $b$ are constant parameters.
These parameters 
have a different physical meaning to parameters in other relativistic scalar field theories, such as the $\phi^4$ model \cite{csw86}.
Their values are not determined by the physics of point particles, but by the properties
of a macroscopic wave-mechanical object with collective degrees of freedom;
which Bose liquids and condensates are \cite{z12eb}.

Correspondingly, 
nonlinear coupling $b$
is a linear function of the 
wave-mechanical temperature $\tps$,
which is defined, in a non-relativistic regime, as 
a thermodynamic conjugate
of 
the Everett-Hirschman's information entropy,
\[ 
S_\Psi = 
-  
\langle \Psi | \ln{(|\Psi|^2/\dnc)}|\Psi \rangle
=
-\int |\Psi|^2 \ln{(|\Psi|^2/\dnc)} \, \dvol
,\] 
where 
$\Psi = \Psi (\vec x, t)$ is condensate wavefunction, 
constant 
$\dnc$ is the decoupling density
(a non-relativistic analogue of $a$, which compensates dimensionality of condensate wavefunction inside the logarithm),
and the integral is taken over the volume occupied by the liquid,
further details can be found in \Ref \cite{az11}. 

Thus, $b$ is not an \textit{a priori} assumed parameter of the model, but a value related to the wave-mechanical thermodynamical properties of the system and its environment 
\be\lb{btemp}
b \sim \tps - \tps^{(0)} 
, 
\ee
where $\tps^{(0)}$ is a reference or critical value \cite{z18zna,z18epl,z19ijmpb}.

Since the potential \eqref{epot}
is defined up to an overall sign, we choose this sign so that, under our signature and curvature conventions,
the scalar field equation would have a static Gaussian solution in the Minkowski spacetime limit \cite{ros68}.
This would correspond to selecting the non-topological soliton sector of the model; a detailed discussion of topological structure can be found in \Refs \cite{z11appb,z18epl,z19ijmpb}.
We  assume that
$b > 0$ in what follows.



Furthermore, 
field equations can be derived from the Lagrangian \eqref{elagr}
in a standard way:
\be\lb{eefe}
R^\mu_{\nu} - \frac{1}{2} \delta^\mu_{\nu} R 
= 
8 \pi G T^\mu_{\nu} 
	,
\ee
where the energy-momentum tensor is determined by complex scalar field:
\ba
T^\mu_\nu 
&=& 
\frac{1}{2}g^{\mu\sigma}
\left(
\nabla_{\sigma}\phi^*\nabla_{\nu}\phi + \nabla_{\sigma}\phi\nabla_{\nu}\phi^* 
\right)
\nn\\&& \quad
-
\frac{1}{2}
\delta^\mu_\nu
\left(
g^{\alpha\beta}\nabla_{\alpha}\phi^*\,\nabla_{\beta}\phi
+
2
\lanp (\phi,\phi^*)
\right)
.
\lb{e:EMT}
\ea
The complex scalar field equation can be extracted from
these equations via Bianchi identities, or by varying Lagrangian \eqref{elagr}
with respect to scalar field, 
and written in a form
\be\lb{esfgen}
\nabla_\mu\nabla^\mu \phi = 2 
\frac{
\partial 
}
{\partial \phi^*} 
\lanp
(\phi,\phi^*)
=
- 2 b 
\ln{(|\phi|^2/a)}
\,
\phi 
.
\ee
This equation is a Lorentz-covariant analogue of the logarithmic Schr\"odinger equation,
which was extensively studied:
to mention only a few very recent works \cite{aff19,wz19,bcs19,ct19,lzh19,ss20}.
Relativistic wave equations with logarithmic nonlinearity have also been extensively studied in the past,
assuming fixed Minkowski spacetime
\cite{ros68,z11appb,dz11,z10gc,szm16}.

Let us consider now spherically symmetric and time independent
solutions of Einstein field equations \eqref{eefe}.
The line element can be written 
in static coordinates 
\be\label{emetric}
	ds^2 = - B(r) dt^2 + A(r) dr^2 + r^2\left(
    d\theta^2 + \sin^2 \theta d \varphi^2
    \right) 
,
\ee
and,
due to a symmetry of the problem, we assume our scalar to be spherically symmetric and stationary
\be\lb{ephi}
	\phi (r,t) = e^{-i \omega t} \Phi(r) 
,
\ee
where $\Phi(r) $ is a real-valued function.
Correspondingly,
field equations \eqref{eefe} and \eqref{esfgen} 
reduce to a system of three ordinary differential equations
\be
\frac{A'}{A^2 x} 
+
\frac{A-1}{A x^2}
= 2
\left[\frac{\Omega^2}{B} + 1 - \ln (\sigma ^2/k)
\right]\sigma ^2 
+ \frac{ (\sigma')^{2}}{A}
,
\lb{eeqsA} 
\ee
\be
\frac{B'}{AB x} 
-
\frac{A -1}{A x^2}
= 
2\left[
\frac{\Omega^2}{B} - 1 + \ln (\sigma ^2/k)
\right]\sigma ^2 + \frac{(\sigma')^2}{A}
,
\lb{eeqsB} 
\ee
\be
\sigma '' + 
\left(
\frac{2}{x} - \frac{A'}{2A} + \frac{B'}{2B}
\right)\sigma' +  
2 A
\left[
\frac{\Omega^2}{B}+
\ln (
\sigma^2/k
)
\right]\sigma 
= 0,
\lb{eeqsSF}
\ee
where  $x = r/L$, 
$L=1/\sqrt{b}$, 
$\sigma=(4 \pi G)^{1/2}\Phi
$, 
$k=4\pi G a$, 
$\Omega
=\omega/\sqrt{2 b}$, and a prime denotes $d/d x$.
If we define
\be
A (x)
=\left[
1- 2 \massf (x)/x
\right]^{-1}
, \ee
we can replace \eq \eqref{eeqsA} with
\be
\massf' (x) =  x^2 
\left\{
\left[\frac{\Omega^2}{B} + 1 - \ln (\sigma^2/k) 
\right]\sigma^2
+\frac{(\sigma')^2}{2A} 
\right\},
\lb{eeqsM}
\ee
and deal 
with a dimensionless mass function $\massf (x)$ in actual computations.

\begin{figure}
\centering
\subfloat[$k< \kspl$]{
  \includegraphics[width=\sct\columnwidth]{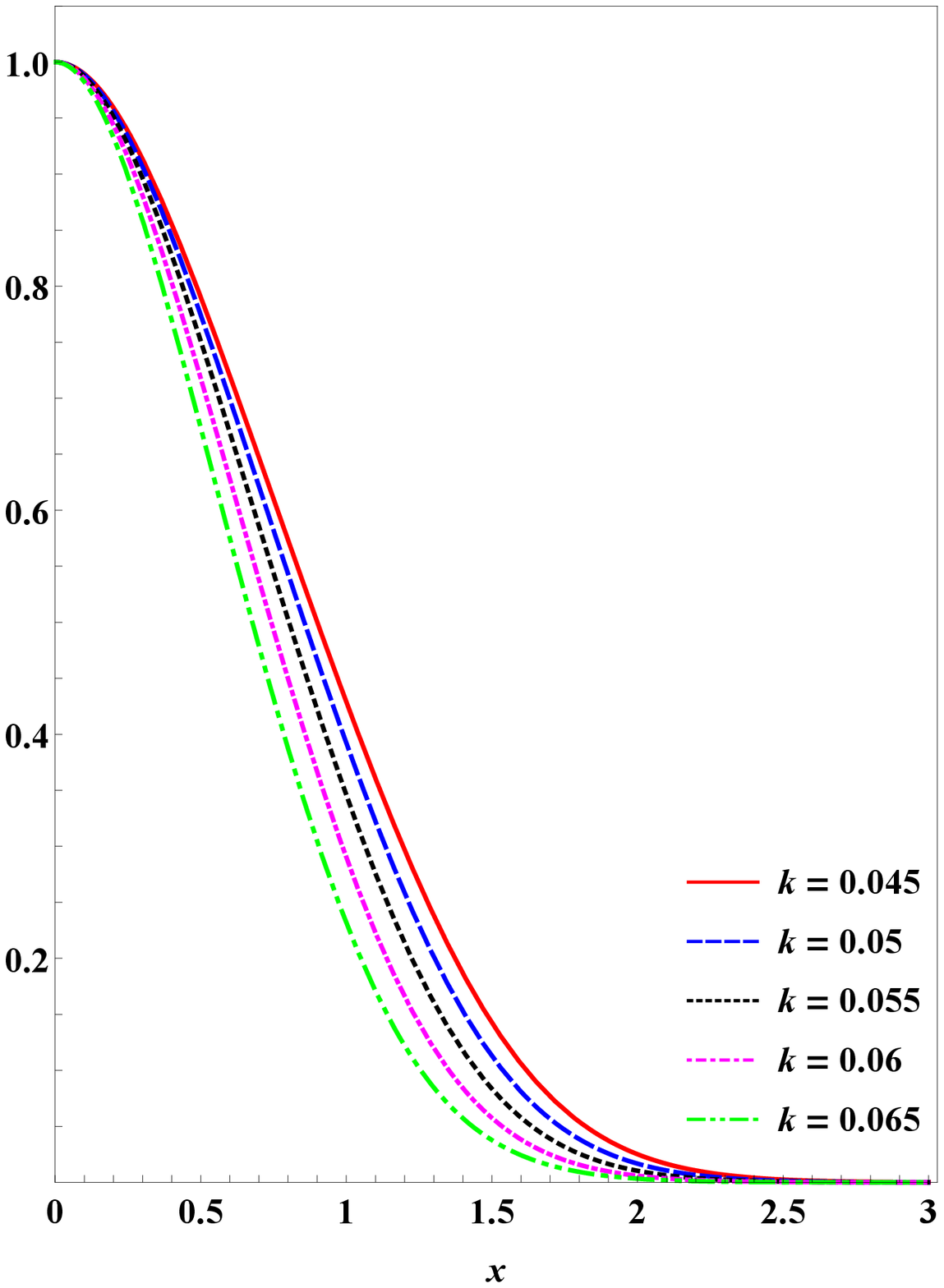}
}
\subfloat[$k> \kspl$]{
  \includegraphics[width=\sct\columnwidth]{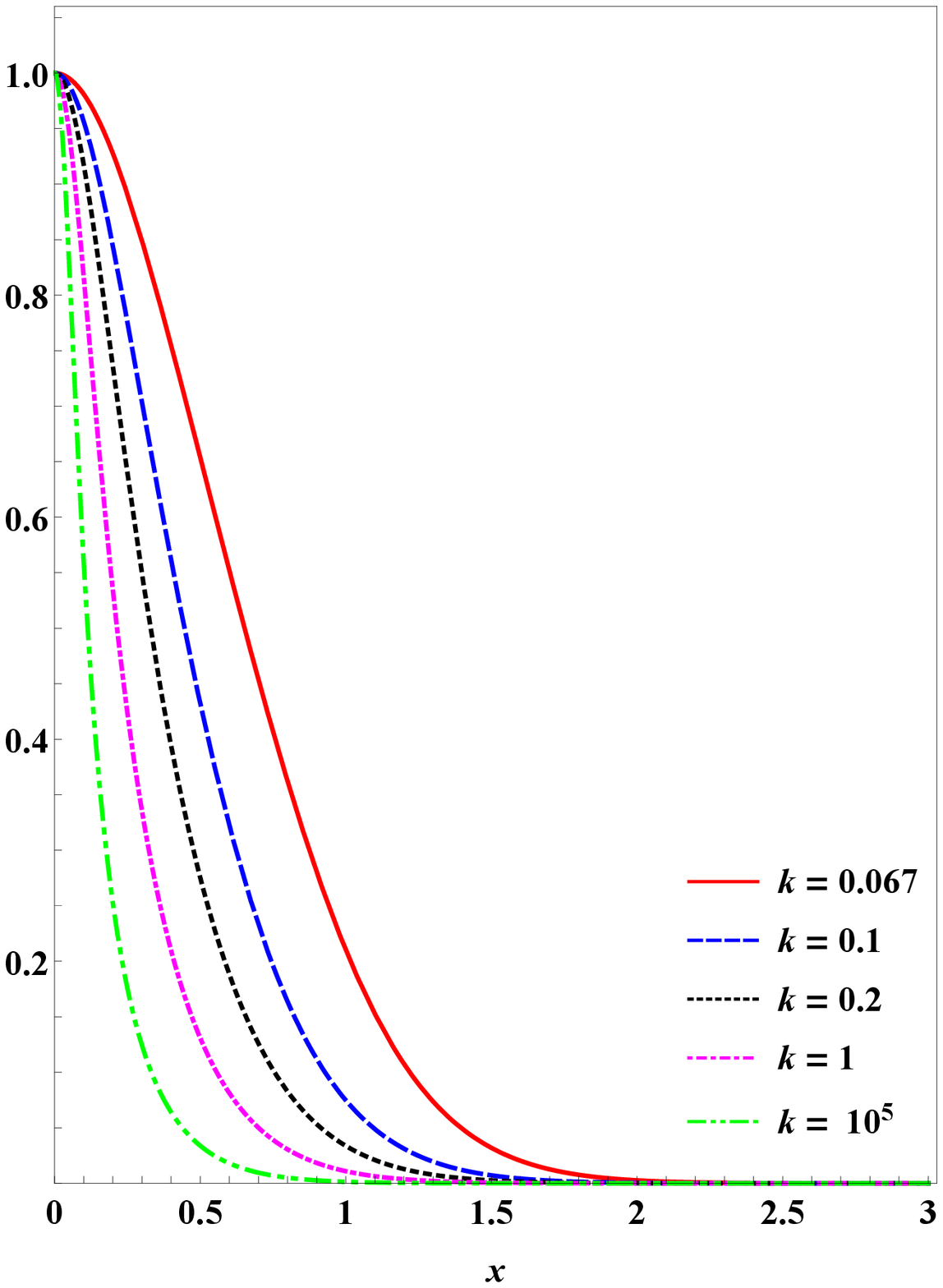}
}
\caption{Function $\sigma (x)$ 
for different values of $k$,
at
$\sigc=1$.
}
\label{sigma-functions}
\end{figure}

\begin{figure}
\centering
\subfloat[$k< \kspl$]{
  \includegraphics[width=\sct\columnwidth]{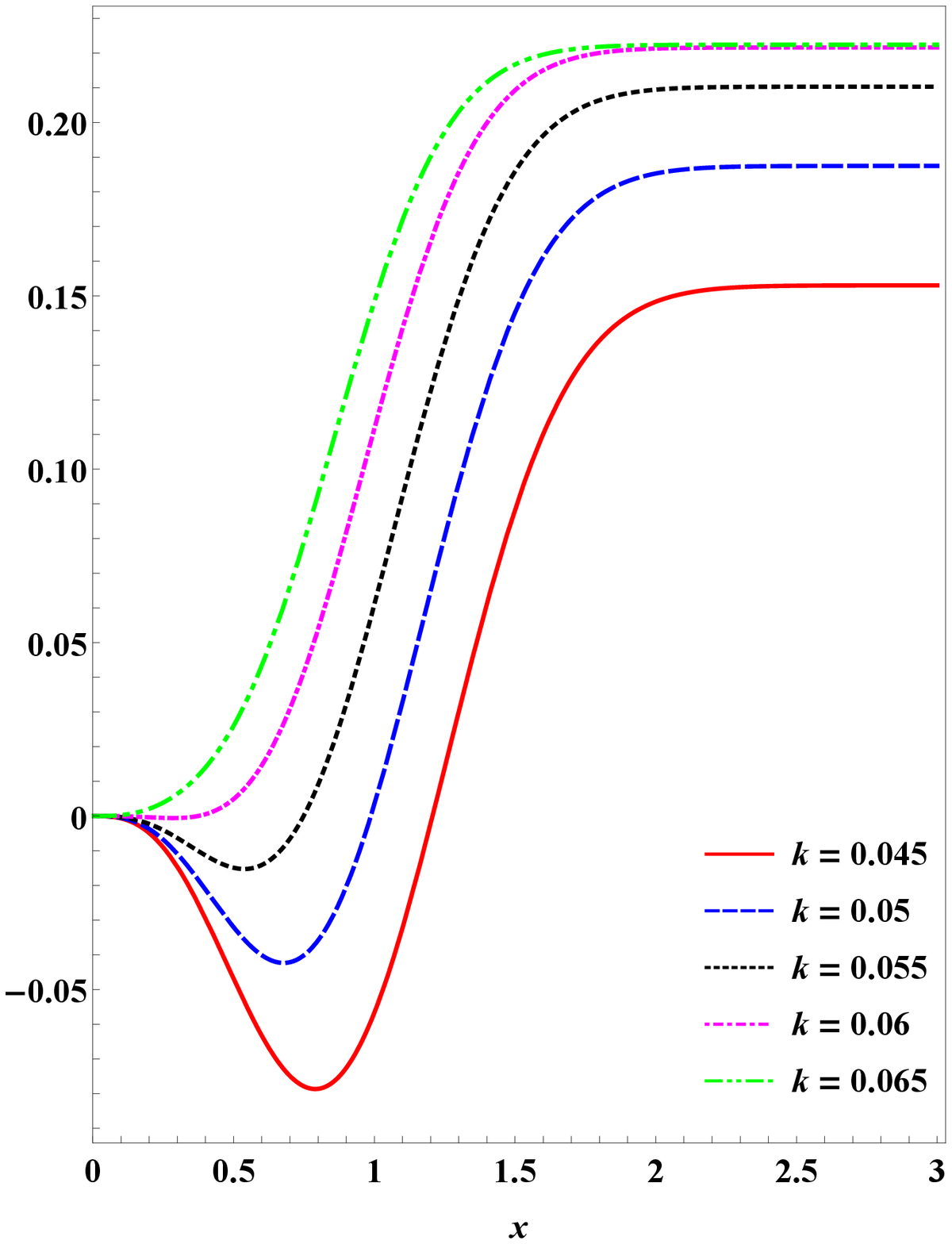}
}
\subfloat[$k> \kspl$]{
  \includegraphics[width=\sctt\columnwidth]{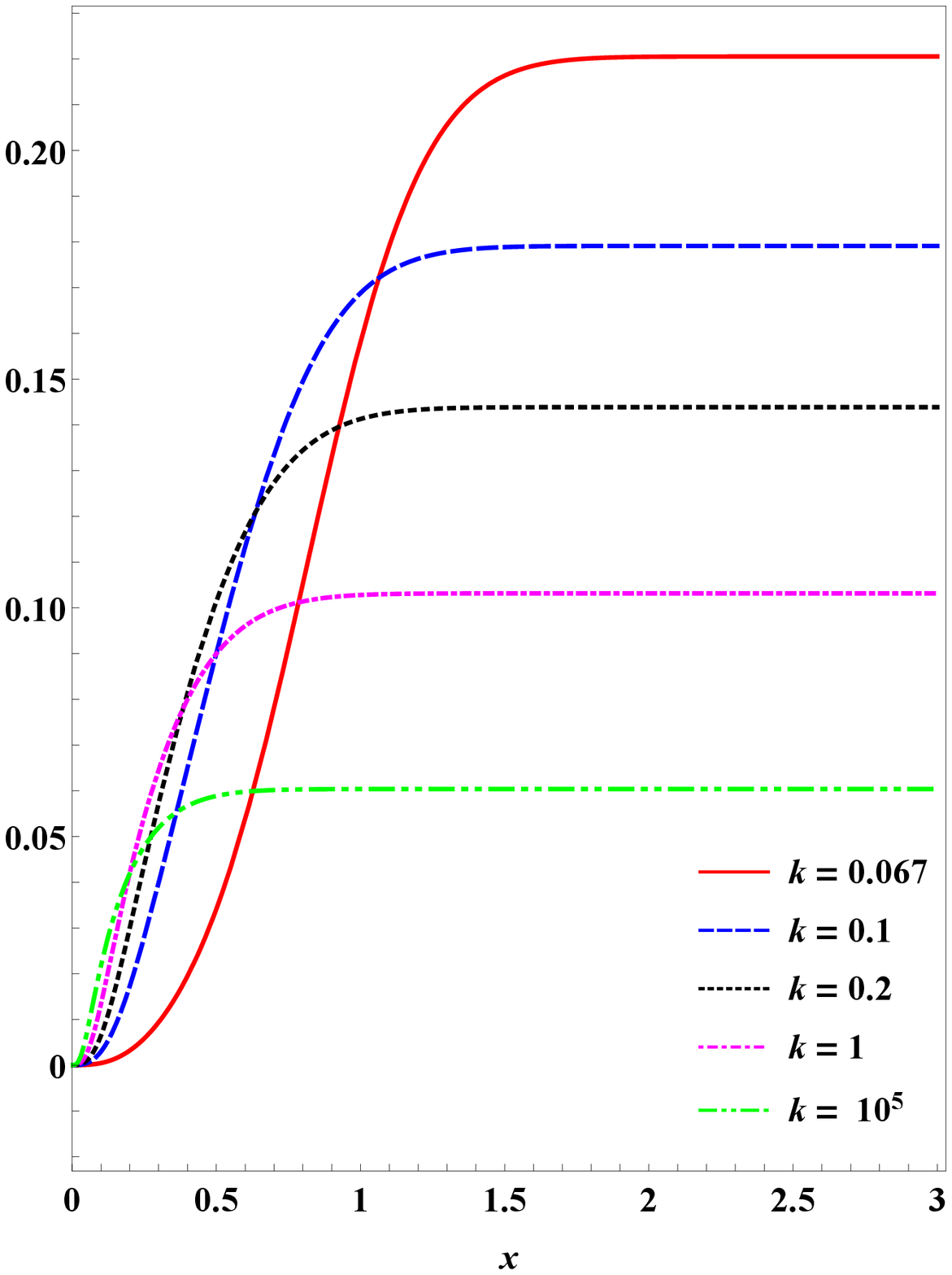}
}
\caption{
Function $\massf (x)$ 
for different values of $k$, at $\sigc=1$.
}
\label{mass-functions}
\end{figure}

\scn{Numerical analysis}{s:numsol}
We numerically solve \eqs \eqref{eeqsA}-\eqref{eeqsM} 
using the shooting method \cite{rb69,csw86}. 
According to this method,
values of parameters 
are 
chosen in such a way that  
our scalar field vanishes at spatial infinity,
and has no nodes and singularities. 
Then, nonsingular finite-mass solutions of \eqs \eqref{eeqsA}-\eqref{eeqsM} are expected to describe the lowest-energy bound states.
Thus,
\eqs \eqref{eeqsA}-\eqref{eeqsM},
together with the
initial conditions
$\sigma(0)=\sigc$, $\sigma'(0)=0$, 
$ \massf (0)=0$, 
and $B(0)=B_0$,
are regarded
as an eigenvalue problem for
$\Omega$ and $B_0$.
Once these are computed,
the total mass can be derived from an asymptotic value of $\massf (x)$:
$M = \massf (\infty) L/G = \massf (\infty)/(G \sqrt b)$.


Computations indicate that: when parameter $k$ goes above a certain critical value $\kspl$, then a 
level splitting occurs
at $\sigc \geqslant \sspl$,
where $\sspl$ is a critical value of the central scalar field.
We have numerically established that
\be
\kspl \approx 0.066
,
\ee
whereas
the value $\sspl$ is different for each $k$.
Therefore, 
in our computations, we restrict ourselves to the region 
$\sigc < \sspl$,
which is sufficient for the purposes of this \art.

A profile of the scalar field
is shown in \fig \ref{sigma-functions},
assuming the central field value to be equal to one.
The field rapidly decays at spatial infinity (in the flat-spacetime limit it would have an exact Gaussian shape), it has no nodes and singular points.

\begin{figure}
\centering
\subfloat[$k< \kspl$]{
  \includegraphics[width=\sct\columnwidth]{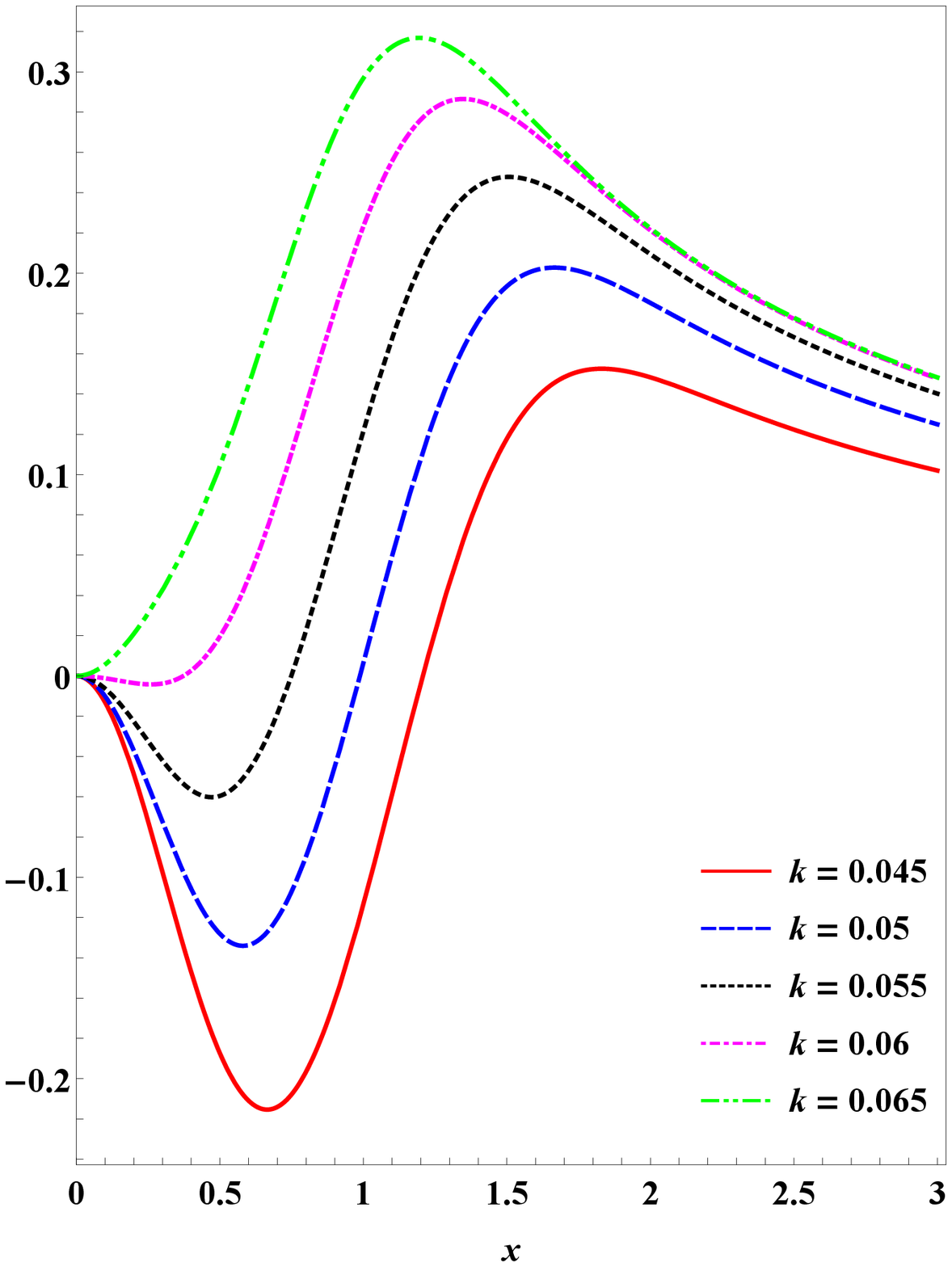}
}
\subfloat[$k> \kspl$]{
  \includegraphics[width=\sctt\columnwidth]{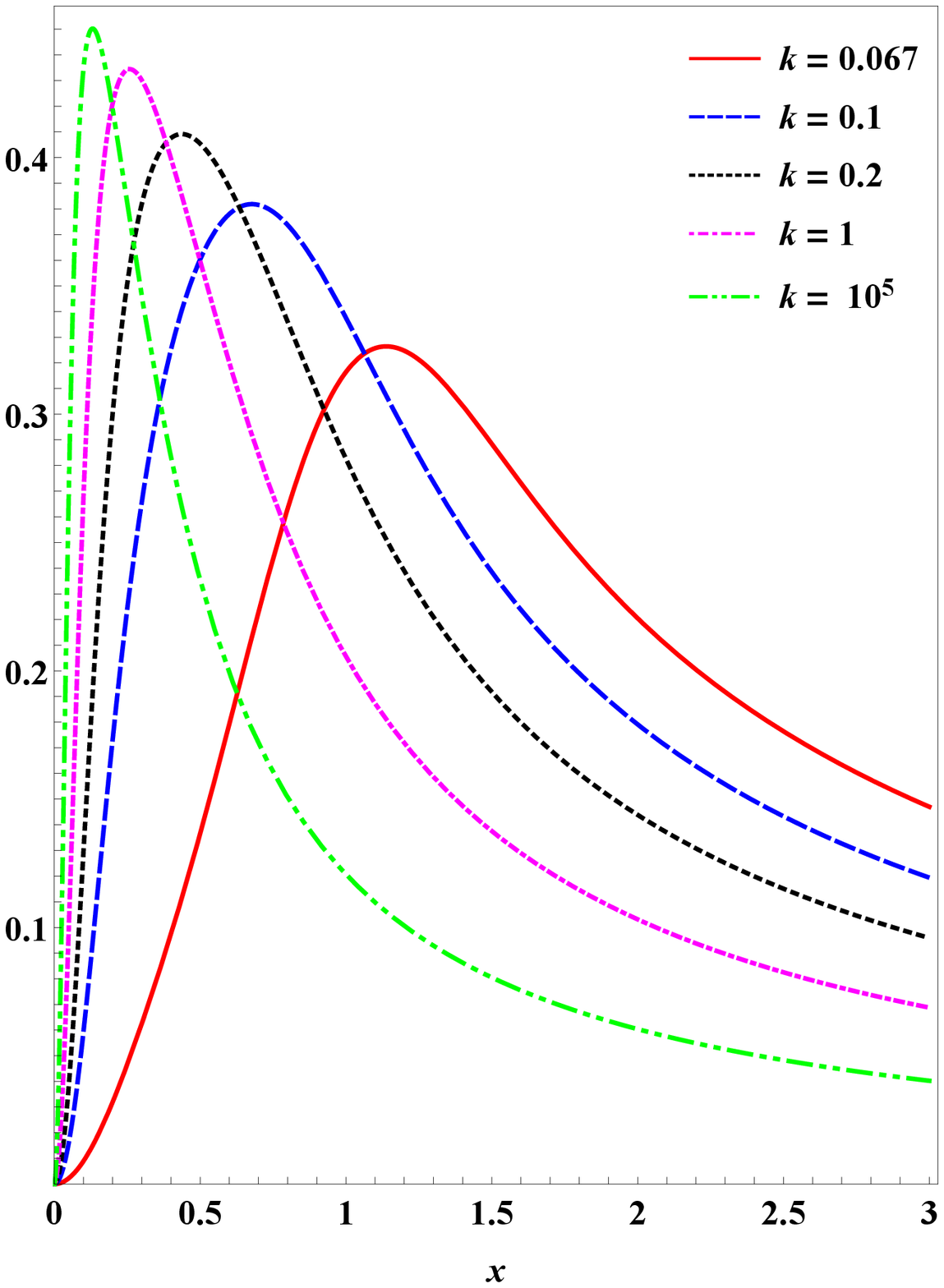}
}
\caption{
Relativisticity function 
for different values of $k$, at  $\sigc=1$.
}
\label{z-functions}
\end{figure}

Profiles of the dimensionless mass function $\massf (x)$ and relativisticity $2 \massf (x)/x$, are shown in \figs \ref{mass-functions} and \ref{z-functions}, respectively.
Unlike the scalar field, their behavior pattern changes as $k$ goes across the critical value $\kspl$.
However, in both cases it indicates that the solutions remain nonsingular, finite-mass and horizon-free.

\begin{figure}[b]
\centering
\subfloat[$k< \kspl$]{
  \includegraphics[width=\sct\columnwidth]{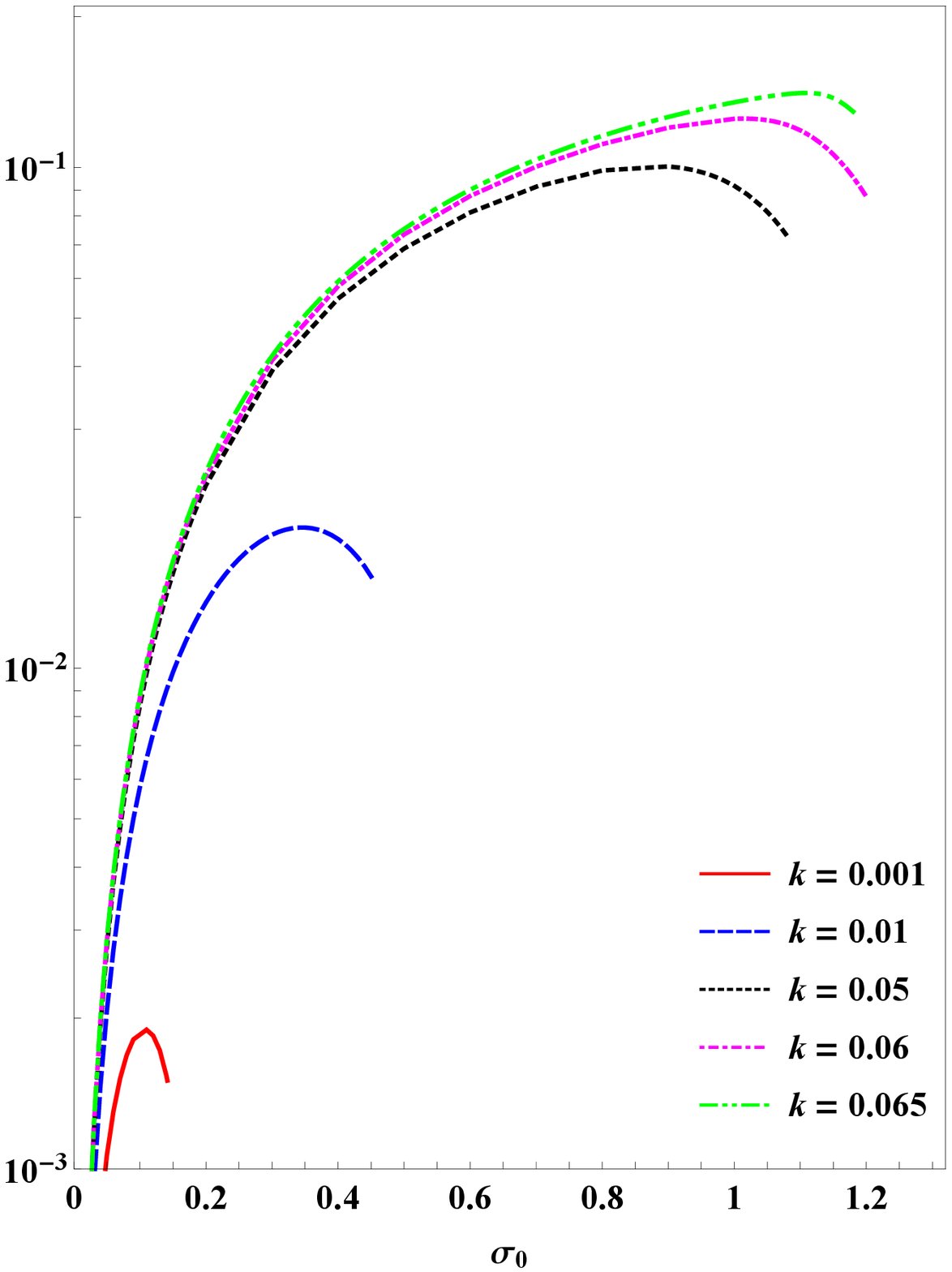}
}
\subfloat[$k> \kspl$]{
  \includegraphics[width=\sct\columnwidth]{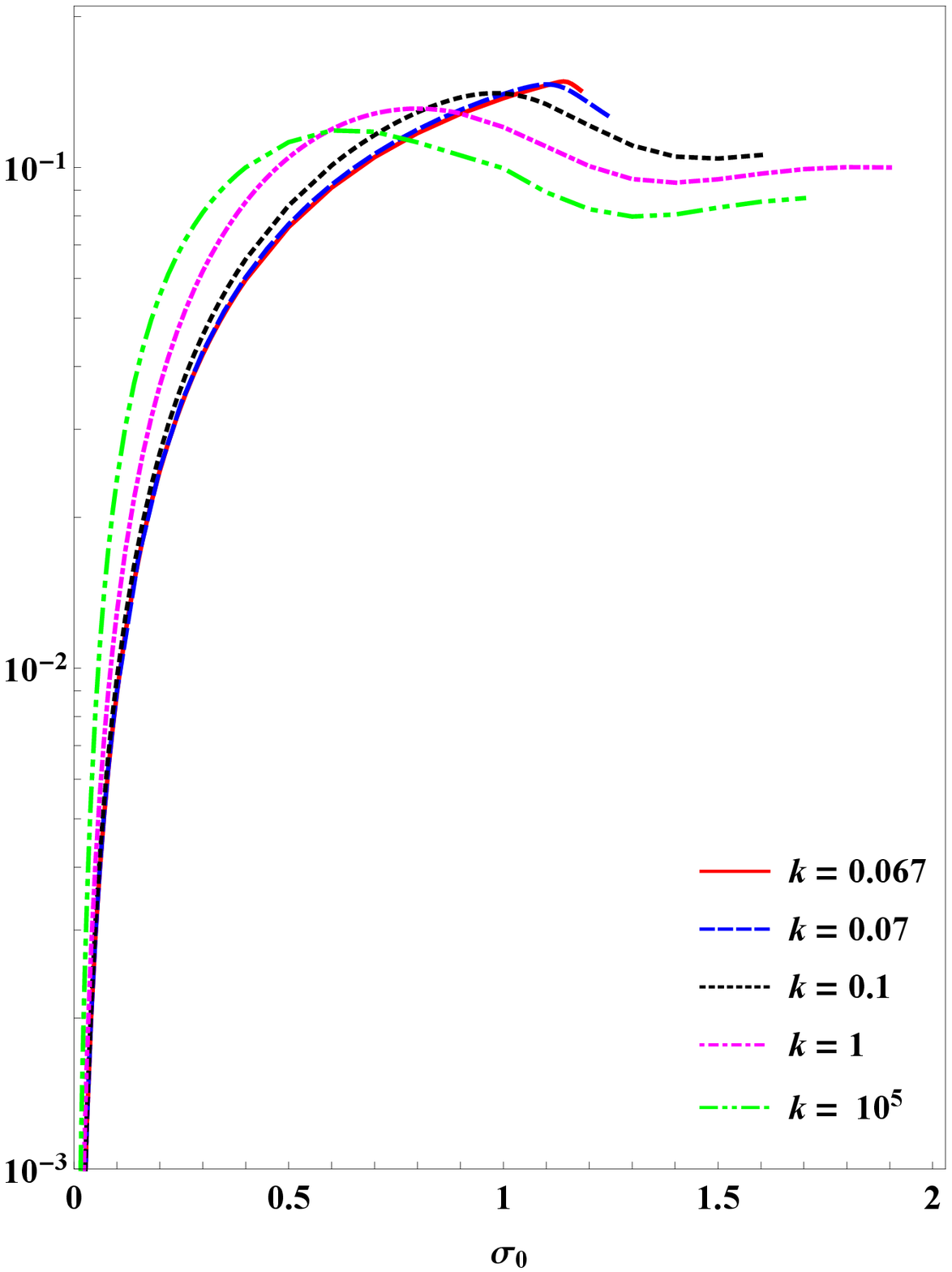}
}
\caption{
Effective compactness versus central field $\sigc$, for different values of $k$.
}
\label{compactness-on-sigma0-at-different-ks}
\end{figure}

The effective compactness of our solution can be defined as the ratio $\massf_{99}/x_{99}$, where 
$\massf_{99} = 0.99 \massf (\infty)$,
and
$x_{99}$ is the dimensionless radius containing $\massf_{99}$.
Its profile, shown in \fig \ref{compactness-on-sigma0-at-different-ks},
is similar to that of boson stars formed by
strongly self-interacting scalar fields \cite{abb10}.
The figure suggests that maximum compactness is achieved in the models in which $k$ approaches the  $\kspl$ value.

From these figures one can estimate that most of the field's energy and star's mass are localized inside the radius 
\be\lb{avrad}
R  
\approx \alpha L \sim b^{-1/2}
,
\ee
where $\alpha$ is a number of order one (its exact value depends on $k$ and $\sigc$).
Therefore, in general relativity,
logarithmic superfluid tends to form lumps whose dimensions scale as $b^{-1/2}$.
What about the bounds on their mass?

\begin{figure}
\centering
\subfloat[$k< \kspl$]{
  \includegraphics[width=\sct\columnwidth]{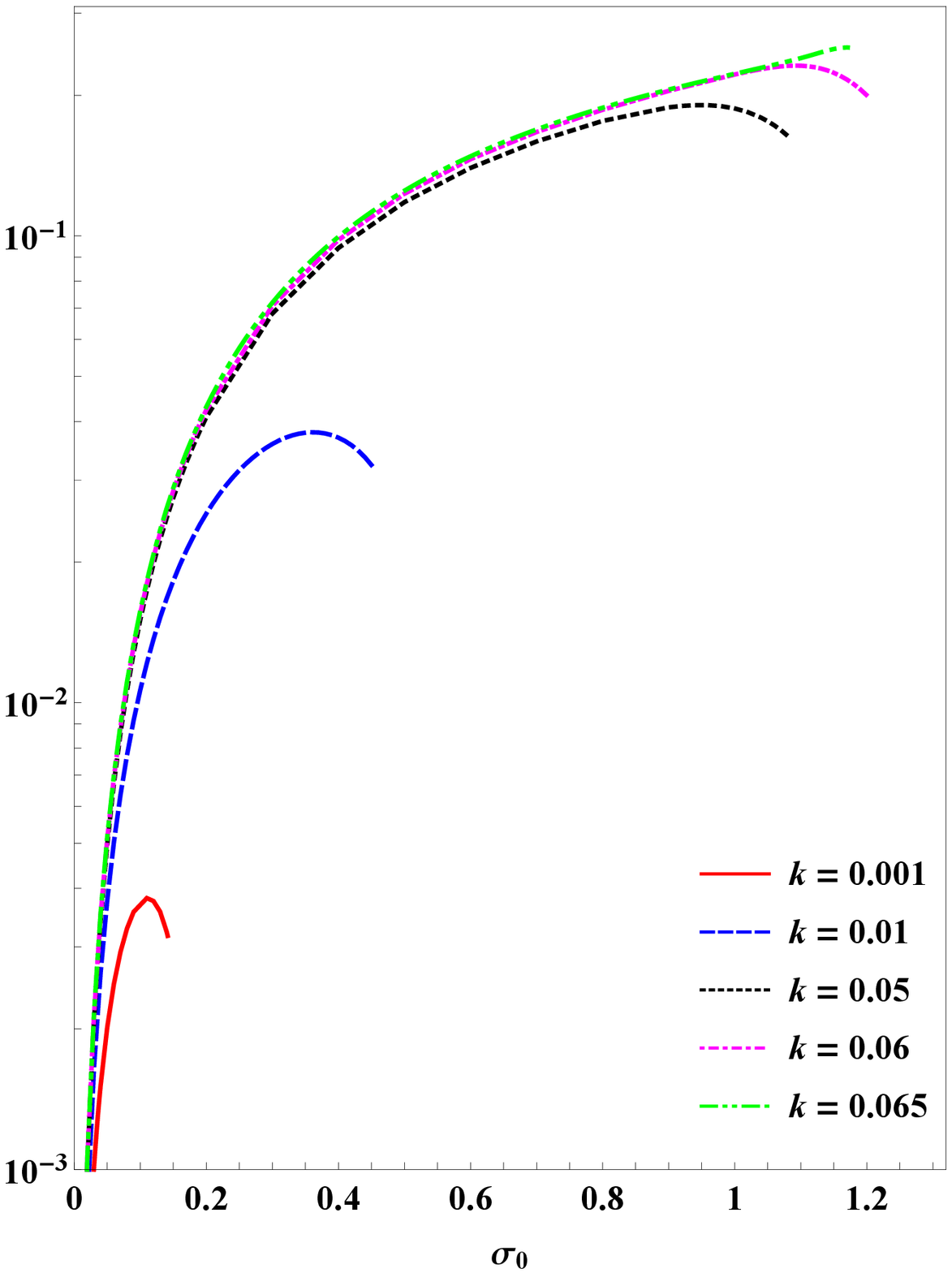}
}
\subfloat[$k> \kspl$]{
  \includegraphics[width=\sct\columnwidth]{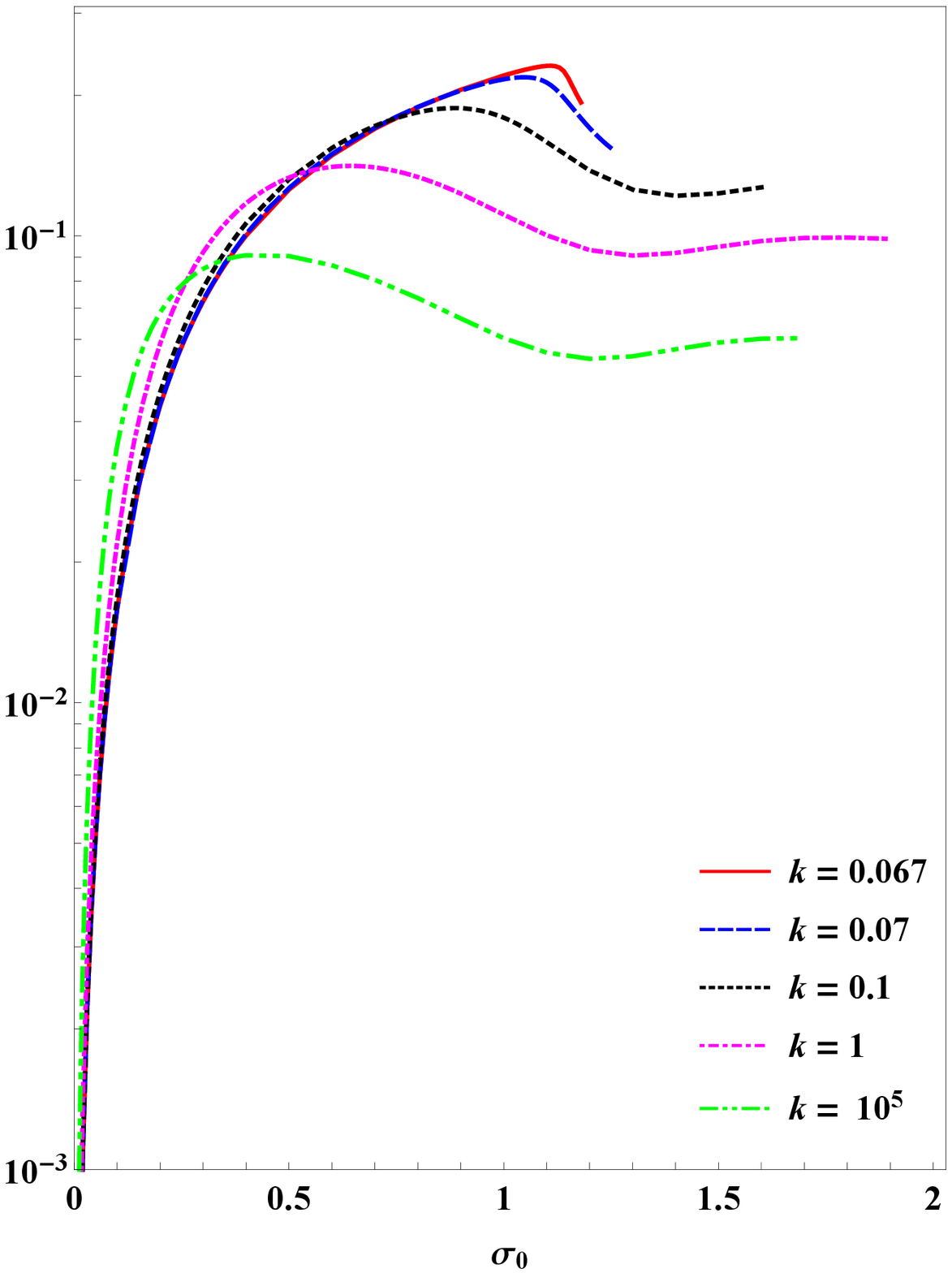}
}
\caption{
Asymptotic value $\massf (\infty)$ 
versus $\sigc$,
for different values of $k$.
}
\label{mass-on-sigma0-at-different-ks}
\end{figure}

\begin{figure}
\centering
\subfloat[$k< \kspl$]{
  \includegraphics[width=\sct\columnwidth]{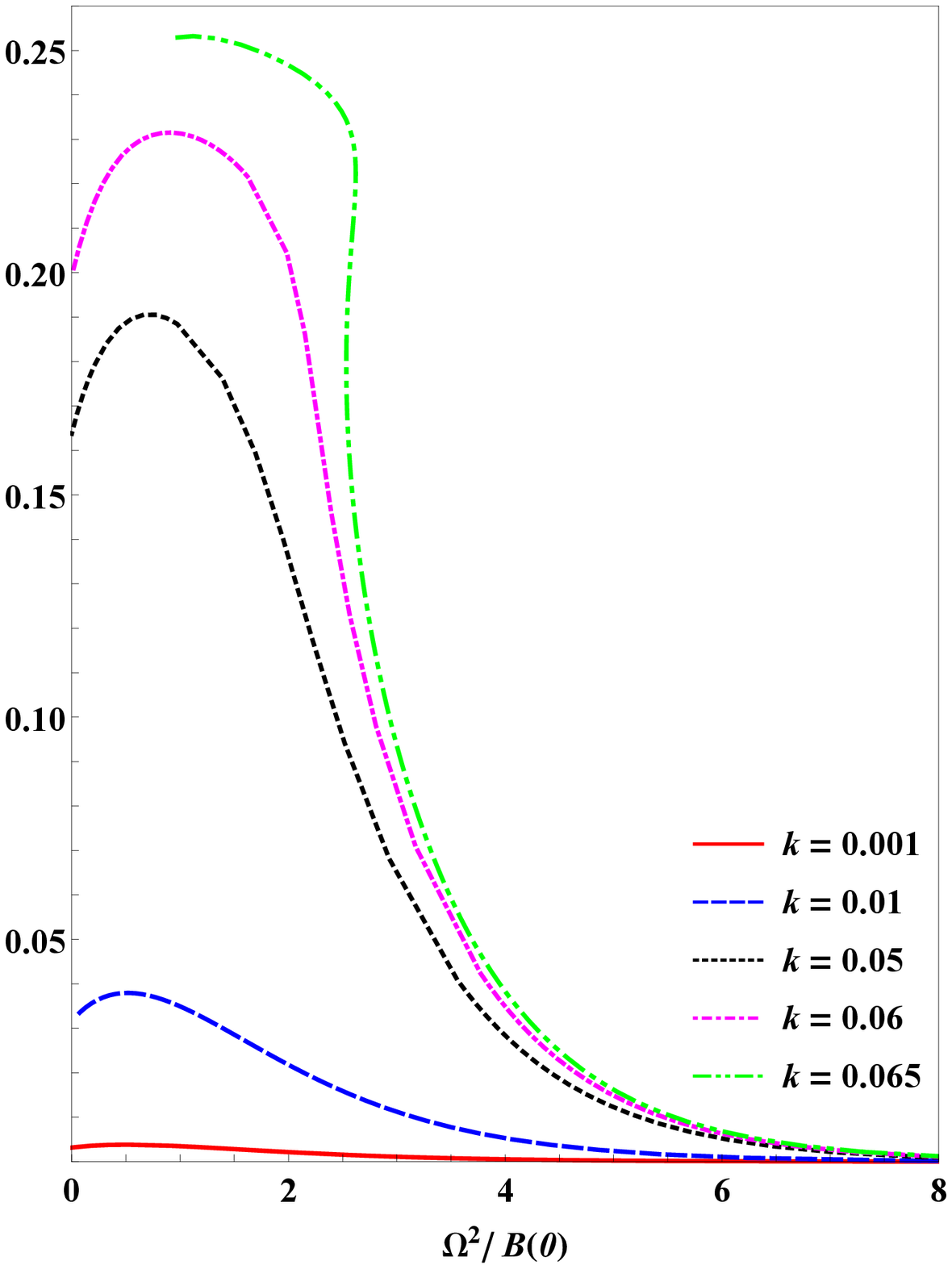}
}
\subfloat[$k> \kspl$]{
  \includegraphics[width=\sct\columnwidth]{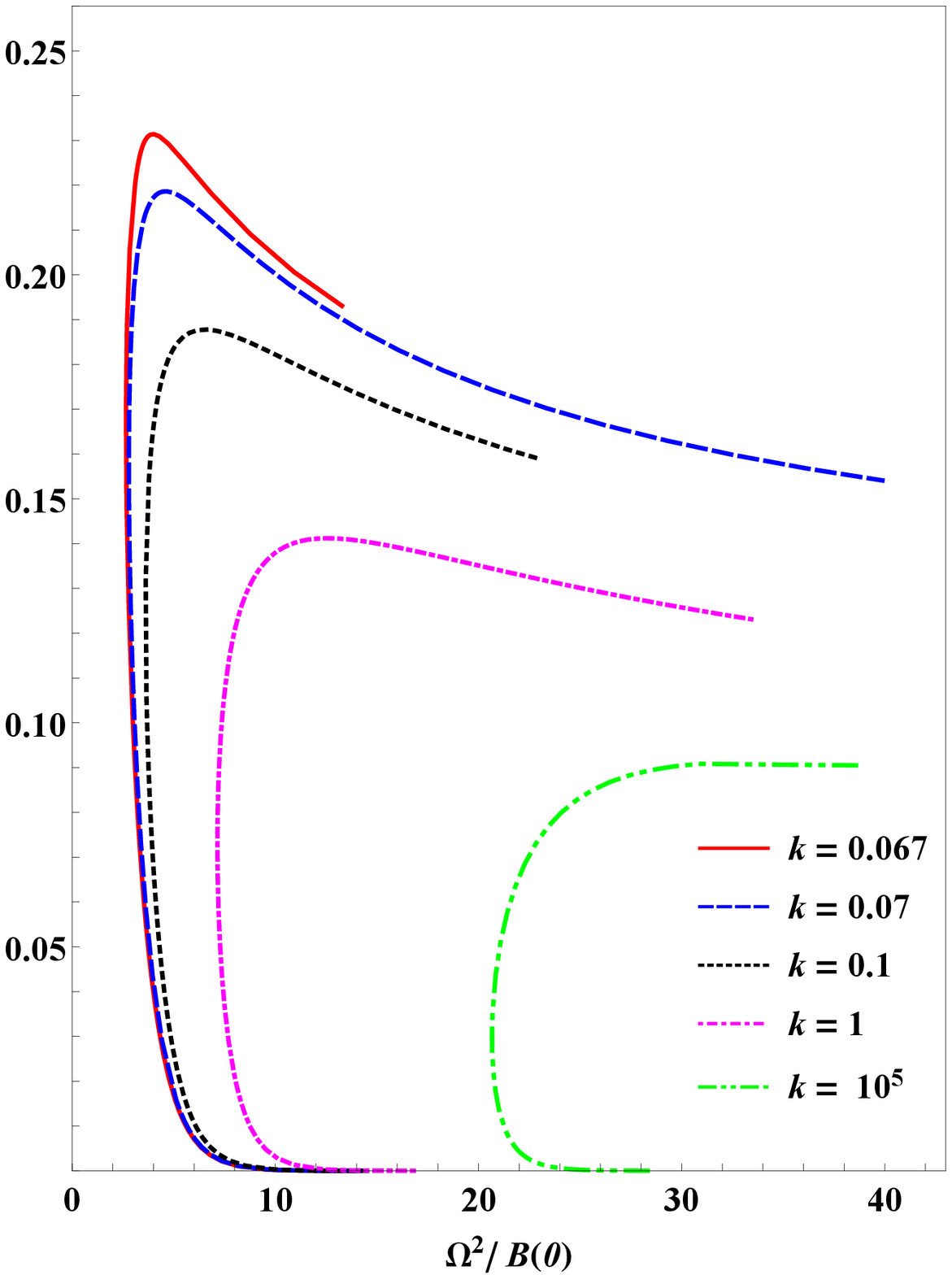}
}
\caption{
Asymptotic value $\massf (\infty)$ 
versus $\Omega^2/B (0)$,
for different values of $k$.
}
\label{mass-on-Omega-sq-over-B0-at-different-ks}
\end{figure}

\begin{figure}
\epsfig{figure= 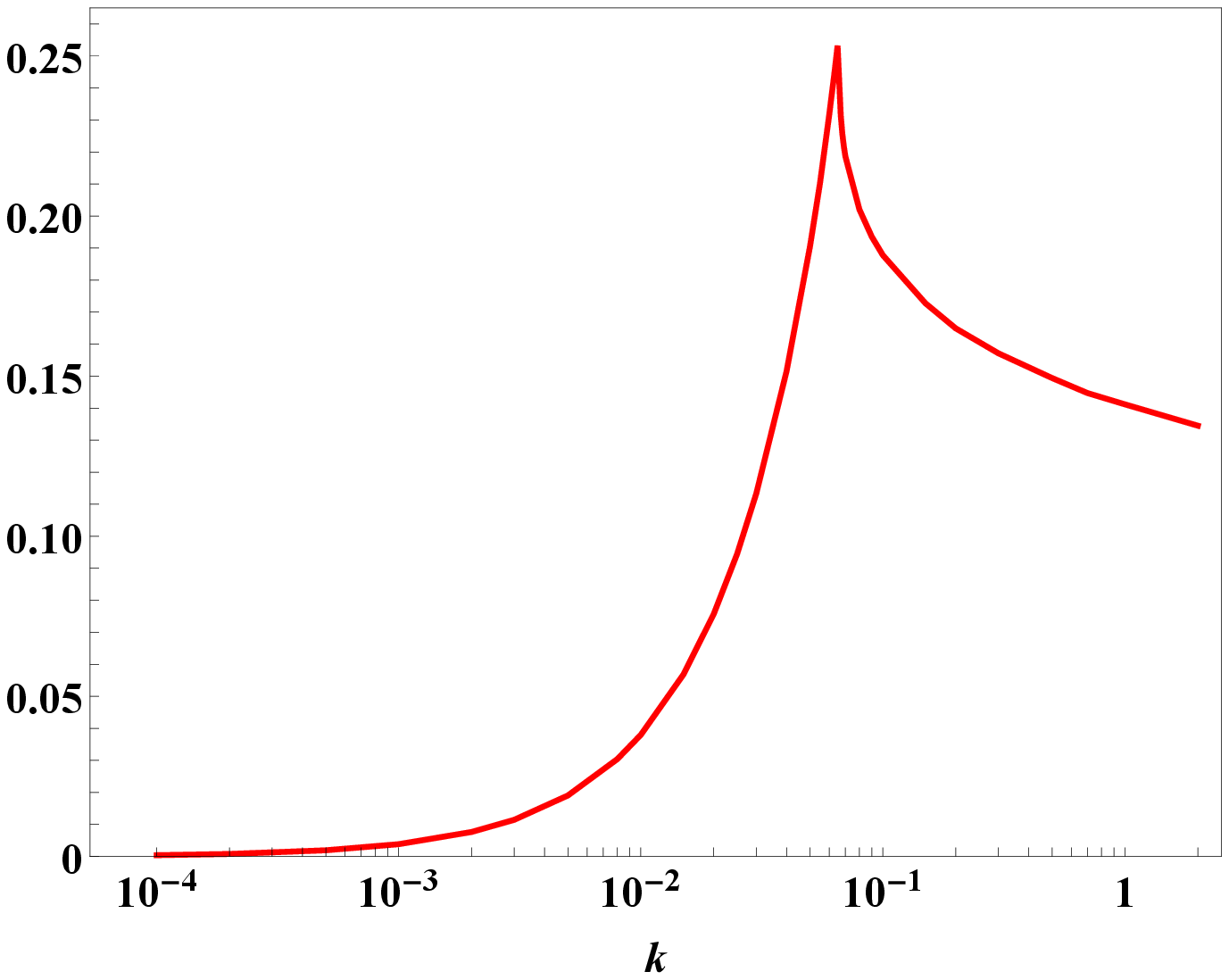,width=  0.96\columnwidth}
\caption{Profile of 
$\massf_\text{max} = \text{max} \left[ \massf (\infty) \right]$ 
versus $ k$.~~~~~~~~~~~~~~~~~\\~
}
\label{max-mass-on-k}
\end{figure}

Profiles of the asymptotic values of the 
asymptotic $\massf (\infty)$  as a function of central field value and frequency are given, respectively, in \figs \ref{mass-on-sigma0-at-different-ks}
and
\ref{mass-on-Omega-sq-over-B0-at-different-ks}.
Furthermore, \fig \ref{max-mass-on-k} is a summarizing plot of the maximum values of $\massf (\infty)$.
One can see that the absolute maximum for $\massf(\infty)$
is achieved at $k \to \kspl$,
which yields the relation
\be\lb{maxmass}
M_\text{max} 
\equiv
\text{max} [\massf (\infty)] L /G 
\approx
(4 G \sqrt b)^{-1}
.
\ee
Therefore, masses of equilibrium configurations of the logarithmic model scale as an inverse square root of nonlinear coupling, similarly to the sizes, cf. \eq \eqref{avrad}.

\scn{Conclusions}{s:con}
It is shown that in general relativity, self-interacting logarithmic scalar fields can form  equilibria, which are described by nonsingular, finite-mass and horizon-free solutions.
According to \eqs \eqref{avrad} and \eqref{maxmass},
their characteristic scales of mass and size  
are determined by the nonlinear coupling:
$M \sim b^{-1/2}$ and $R \sim b^{-1/2}$.
Since this coupling has no known bounds other than being positive-definite, see discussion in
between \eqs \eqref{epot} and \eqref{eefe},
both mass and size of the equilibrium configurations have no upper and lower bounds.
Therefore, our model can be used to describe objects with lengths and masses of a wide scale range,
which agrees with the dilatation symmetry mentioned above.

If $b$ is small, then our model refers to astronomical-scale CS/BHM objects, such as superfluid stars
and cores of neutron stars \cite{mi59,ls01}.
For example,
let us assume that our superfluid star has the maximum mass which is equal to the mass of Sun,
$
M_\odot \approx 2\times 10^{33}\, \text{g}
$.
Then from \eq \eqref{maxmass}, we obtain  
\[ 
b_\odot 
\equiv
b ({M_\text{max} = M_\odot}) 
=
(4 G M_\odot)^{-2}
\approx
2.8 \times 10^{-12}\, \text{cm}^{-2}
.\]
Correspondingly, \eq \eqref{avrad} gives
us an estimate for the radius of such a star:
$ 
R_{M_\odot} 
=
\alpha/\sqrt{b_\odot} 
\approx
\alpha (2.8 \times 10^{-12})^{-1/2} \, \text{cm}
\sim
\alpha (6 \times 10^{5}) \, \text{cm}
\lesssim
10 \, \text{km}
$. 

When $b$ is large, then the model describes composite particlelike objects of finite size, such as Q-balls, which self-interact and curve spacetime
(logarithmic Q-balls in fixed flat spacetime were studied in \Refs \cite{ros68,z11appb,dz11,mo20}).

Furthermore, if we assume wave-mechanical temperature to be proportional to the thermal one, $\tps \propto T$,
then \eqs \eqref{btemp} and \eqref{maxmass} suggest that mass of logarithmic superfluid stars and Q-balls scales 
as an inverse square of temperature, 
$
M \sim (T - T_c)^{-1/2}
$,
where $T_c \propto \tps^{(0)}$.
It means that temperature of these objects should depend on their mass as
\be
T \sim M^{-2}
, 
\ee
up to an additive constant $T_c$.
If this constant is close to absolute zero, then massive superfluid stars alone 
must be cold objects with low evaporation rate by thermal radiation,
whereas logarithmic Q-balls have a longer lifetime in thermally hot and dense environments.



\begin{acknowledgments}
This research is supported by Department of Higher Education and Training
of South Africa
and in part by National Research Foundation of South Africa.
Proofreading of the manuscript by P. Stannard is greatly appreciated.
\end{acknowledgments}


\end{document}